# The USNO rubidium fountains

**Steven Peil, James Hanssen, Thomas B. Swanson, Jennifer Taylor, and Christopher R. Ekstrom**

United States Naval Observatory, 3450 Massachusetts Ave. NW, Washington, DC 20392 USA

steven.peil@usno.navy.mil

**Abstract**. Four rubidium fountains at the U.S. Naval Observatory (USNO) have been in operation for 4.5 years. Each fountain has demonstrated periods of stability marked by Total or Theo deviation below $10^{-16}$. Occasional frequency changes, on order of 1.5 times per year per fountain, introduce deviations from white-frequency noise behavior. Averaged together, the four fountains form an ensemble with a white-frequency noise level of $10^{-13}$ and excellent long-term stability as compared to the primary frequency standards contributing to TAI. Progress on using the clocks at USNO for improving limits on coupling of fundamental constants to gravity by measuring the universality of the gravitational redshift for different types of clocks is discussed.

## 1. Introduction

Primary frequency standards based on atomic fountains were introduced in the mid-1990s. These laser cooled systems marked a major technological advance over the thermal beam standards of the time, reaching about 50 times better accuracy. On the other hand, thermal-beam systems, along with hydrogen masers, are still the predominant technology for continuously running clocks used for timescale generation. When laser cooled frequency standards first started to show promise, the U.S. Naval Observatory (USNO) initiated a program to develop continuously running atomic fountain clocks to supplement its ensemble of commercial cesium beams and hydrogen masers for precise timing. This program has culminated in six operational rubidium fountain clocks, designated NRF2 through NRF7, at two locations – NRF2, NRF3, NRF4 and NRF5 are at USNO in Washington, DC, and NRF6 and NRF7 are at the Alternate Master Clock facility in Colorado. The ensemble of four fountains in Washington, DC has been declared fully operational and has been running for 4.5 years, and each of the four fountains has been contributing to EAL (Echelle Atomique Libre) without interruption for the past 3.5 years. The results, analysis and discussion presented here focus on these four clocks.

## 2. Rubidium Fountain Overview

The design of the USNO fountains has been discussed in detail elsewhere [1-3]. The systems use rubidium, which has a much smaller cold-collision shift than cesium, trapped directly from background vapor into a magneto-optical trap. External cavity diode lasers (ECDLs) provide the laser light, and compact, mechanically stable optical components are used to ensure long-term optical alignment. The fountains were built two at a time, using a revised design after the first two, NRF2 and NRF3, were built. Each fountain has a 5MHz output that is sent to a clock measurement system, which measures individual clocks against the USNO master clock (MC). The 5MHz signal is generated from a high resolution frequency synthesizer (AOG), which is steered roughly every 20s to the fountain frequency (with a gain that results in a loop time constant of 1 minute). A quartz crystal oscillator that is phase locked to a hydrogen maser provides the reference for both the AOG and for the 6.8GHz drive applied to the microwave cavity. The 6.8GHz drive is steered along with the AOG to keep it centered on the central Ramsey fringe. The steering configuration is illustrated in Fig. 1.

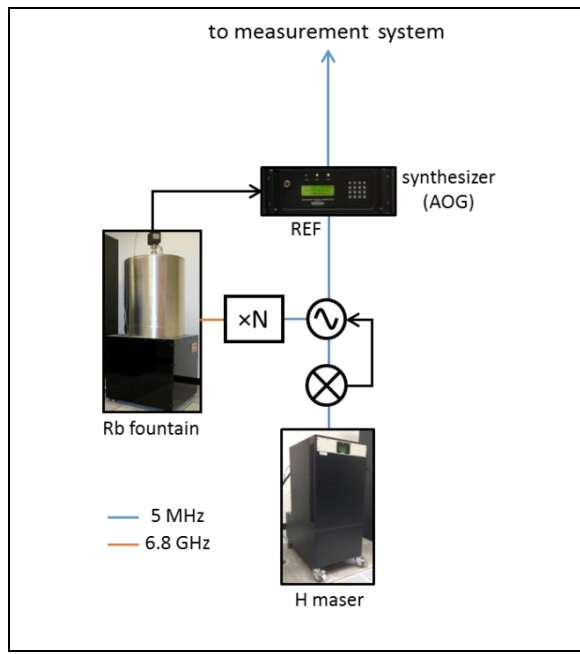

Fig. 1. (colour online). Illustration showing how a rubidium fountain is integrated into our clock measurement system. A 5MHz drive referenced to a hydrogen maser serves as the reference for a synthesizer (AOG) and for the microwave drive for the atoms. The fountain steers the AOG (and the microwave drive applied to the cavity) with a time constant of 1 minute, creating a continuous signal that is measured and compared to other clocks.

Each fountain has its own dedicated maser. Steering of the AOG is suspended if there is an interruption in normal operation, and the system relies on the frequency stability of the maser. Alarms are triggered that indicate attention is required, and steering automatically resumes when the problem is resolved. This graceful holdover is also implemented when scheduled maintenance is carried out, avoiding an interruption in the 5MHz output during remediation. A maintenance schedule has been established which calls for replacement of the ECDLs and the probe beam shutter every two years; the tapered amplifier chips age more slowly and can be replaced less frequently. User intervention can also be precipitated by processor errors and optical misalignment. Dedicated masers, stable optical components, and a well-regulated environment all contribute to an average uptime of greater than 99.5% for the four fountains.

### 3. Rubidium Fountain Performance

*3.1. Individual fountains*
The average short-term stability of the four fountains over 4.5 years of operation is $2\times10^{-13}$ at 1s. For certain intervals of time, this white-frequency noise level dictates performance and measured fountain (in)stabilities can reach the mid $10^{-17}$'s; in fact, each fountain has shown performance consistent with white-frequency noise for periods over 1 year.

In Fig. 2 we show the relative phase difference and the frequency stability of NRF4 and NRF5, which have been the most stable of the four clocks. Figure 2(a) shows performance over a particular 1.5-year interval (within the first 2 years of operation) during which the stability plot is consistent with white-frequency noise behavior out to the maximum averaging time of 6 months (for Theo). This performance is consistent with an average (in)stability per fountain below $5\times10^{-17}$. For this data set, the relative drift between the fountains is zero at the level of $6\times10^{-19}$/day. In Fig. 2(b) we show the entire phase record from the first two years of operation. The stability plot in this case shows a deviation from white-frequency noise, with a slope for long averaging times consistent with $\sqrt{\tau}$. The phase record indicates that after about 0.9 years there is a change in the relative frequency of the fountains. If we process the data and remove this frequency change, of size $<3\times10^{-16}$, the resulting frequency record integrates down, showing no signs of deviation from white-frequency noise. Thus, the appearance of non-stationary behavior is consistent with the occurrence of a single change in the relative frequency over 2 years.

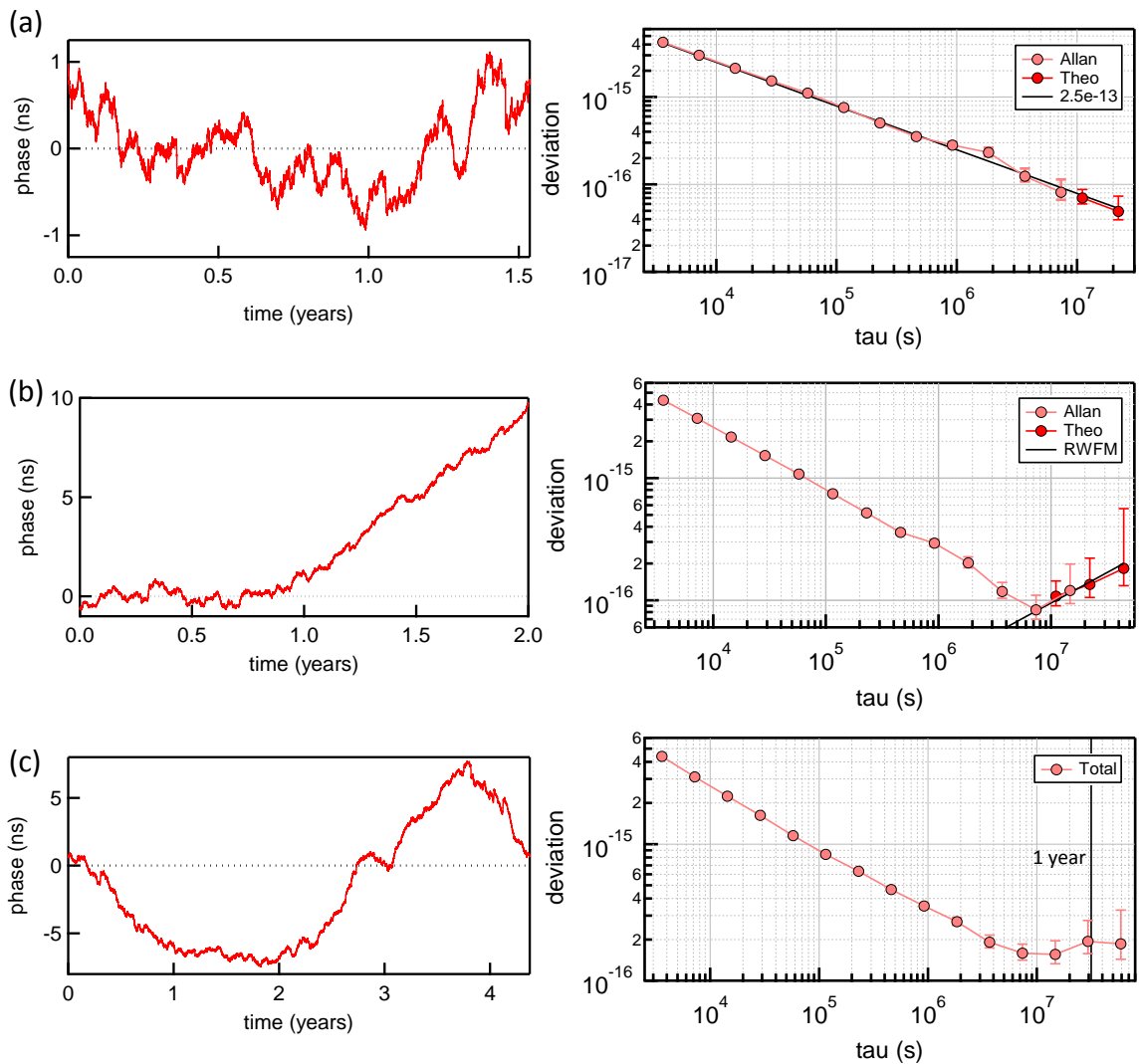

Figure 2. (colour online). Plots of relative phase (left) and stability (right) for NRF4 versus NRF5 for (a) a particular 1.5-year interval, (b) the first two years of operation, and (c) the entire 4.5 years of operation. The value of the frequency removed in each plot is different; the average frequency difference in (a) and (c) are zero and the change in frequency in (b) is emphasized. Solid lines on stability plots are references for (a) $1/\sqrt{\tau}$ and (b) $\sqrt{\tau}$ behavior.

The NRF4 versus NRF5 data for the entire 4.5 years the fountains have been operating are shown in Fig. 2(c). Over this time there have been several changes in the relative frequency, and the result is a stability plot more consistent with flicker frequency noise than with random walk. Even with these frequency changes, the phase record for NRF4 versus NRF5 shows that the fountains remained within 15 ns (±7.5 ns) of each other over 4.5 years.

A three-cornered-hat analysis shows that all of the fountains reaching stabilities below $5\times10^{-16}$, from $4\times10^{-16}$ (NRF3) to $4\times10^{-17}$ (NRF5), for averaging times of a year and longer; see Fig. 3(a). These data points are obtained by averaging the four different ways of calculating individual stabilities from four clocks (using three-cornered-hat analysis). For some of these calculations, negative variances were obtained for the best clocks (NRF4 and NRF5). This failure of the three-cornered hat technique is likely due to disparate stabilities rather than correlations between fountains. Representative error bars are included on the 1-year data points in Fig. 3(a).

*3.2. Frequency changes*

The source of the frequency changes in the fountains is presently unknown. Some seem to correspond with user intervention, but no decisive pattern has yet been determined. Over 4.5 years of operation, there have been on average 1.5 frequency changes per fountain per year, with an average magnitude of $6\times10^{-16}$, but including the sign of the shift, the average value is $6\times10^{-18}$.

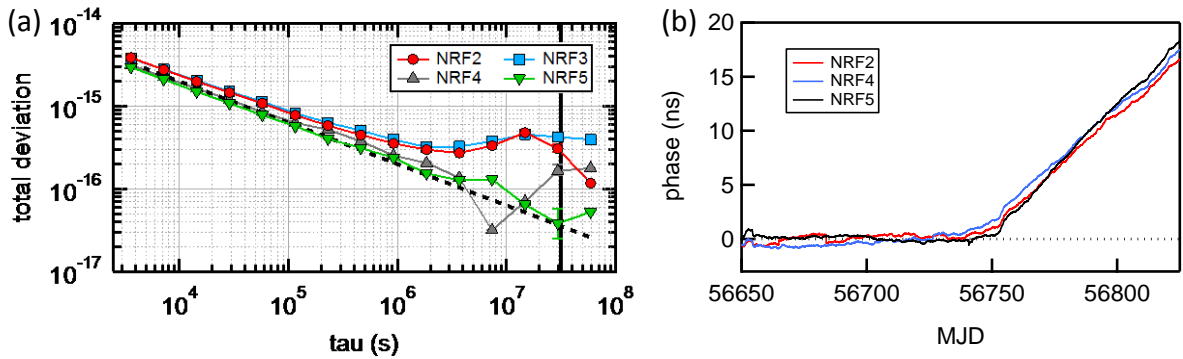

Figure 3. (a) (colour online). Three-cornered hat stability plots for the four fountains, using all of the 4.5 years of data. (b) Example of a frequency shift in NRF3, identified in plots of the relative phase of NRF3 compared to the other 3 fountains.

Obvious candidates for these frequency shifts are changes in the values of systematic biases. The largest drivers of frequency biases are the second-order Zeeman shift, the blackbody shift, and the gravitational redshift. Any changes in the gravitational redshift would be common to all of the fountains and therefore would not appear in frequency comparisons. Similarly, changes in room temperature would be common mode for pairs of fountains, which would yield a unique signature for the frequency changes that is not observed. Additionally, both temperature and humidity are continually monitored and show no changes that are correlated with fountain frequency changes.

We can also rule out the second-order Zeeman term as the source of our frequency shifts. We regularly measure the average magnetic field experienced by the atoms over their trajectory by measuring the field-sensitive transition frequency. We see no changes in magnetic field that correlate in size or time with the frequency changes observed in the fountains.

Another possible cause of frequency shifts that we have studied is the light shift. Laser light is fiber coupled to an optical table with beams splitters, acousto-optic modulators (AOMs), and shutters for the trapping and detection beams. Light is then fiber coupled from this table and delivered to the vacuum chamber. When the shutters block the laser light, scattered light that could conceivably couple to the vacuum chamber is still present. We have added an overall shutter at the output of the laser that prevents any light from reaching this optical table. The use of this additional shutter does not affect the fountain frequency at an observable level, and we therefore do not believe that the fountains are affected by light shifts.

Changes in the size of the cold collision shift over time are calculated to be far too small to be noticeable. Furthermore, changes in the density of the atoms involved in the fountain cycle should be reflected by changes in the number of atoms. We measure this number on each cycle and have not observed a correlation between a change in the number of atoms and a change in the frequency of a fountain.

Systematic biases that involve the microwave cavity and microwave drive are difficult to decouple. We investigated biases associated with the microwave cavity in our engineering prototype, NRF1, but did not carry out detailed measurements in the operational systems. Based on the size of cavity-related biases reported in primary frequency standards, including the distributed cavity phase shift, cavity pulling, and microwave leakage, it is unlikely that any of these are the source of the largest frequency shifts we have seen, though we cannot rule them out as contributing to the smaller shifts.

It is possible that subtle perturbations to the microwave drive are causing the frequency shifts we observe. During laboratory testing, we observed a coupling between RF signals going to the AOMs and the microwave drive going to the cavity in a particular fountain. Additional isolation reduced this coupling beyond our ability to observe; but completely eliminating it at the level we are concerned with is difficult to confirm. We remain suspicious that stray coupling to the microwave drive could be contributing to the shifts that we observe, and we hope to incorporate a microwave drive monitor in the future to investigate this possibility.

*3.3. Fountain timescale*

The primary motivation for building multiple fountains was to maximize the probability that at least one fountain was working at any given time. With the unexpected robustness in fountain operation that has been achieved, marked by an average uptime greater than 99.5%, the existence of four fountains allows us to mitigate the effects of the frequency shifts we observe. Intercomparisons of the relative phases of the four systems typically enables a frequency change to be unambiguously associated with a particular fountain, as illustrated in Fig. 3(b). A post-processed timescale can be created in which the effect of the frequency change is removed. This is a recharacterization of the rubidium fountains analogous to that applied to hydrogen masers when making a timescale. This rubidium fountain-only timescale should demonstrate better long-term stability than the individual clocks. Perhaps the ultimate reference for determining long-term frequency stability is the collection of primary frequency standards that calibrate TAI, and we have demonstrated in the past that our fountain timescale shows no drift with respect to the primary standards at a level of $1.2 \times 10^{-18}$ per day [4]. The average short-term stability of the fountain timescale is $10^{-13}$ at 1 second, improving over the timescale derived from USNO's 70 commercial cesiums by a factor of 10, which translates to a factor of 100 improvement in the averaging time required to reach a particular level of stability.

## 4. Tests of Relativity

The behavior of the frequencies of atomic clocks in time and space has been used to constrain violations of the Einstein Equivalence Principle (EEP), which lays the foundation for General Relativity. One test of EEP includes looking for a difference in the gravitational redshift of clocks of different species as they experience a changing gravitational potential, typically the varying solar potential due to the Earth's elliptical orbit [5]. From the dependence of clock frequencies on fundamental constants, the redshift tests can be used to constrain the dependence of fundamental constants on gravity [6]. While optical clocks provide significantly higher precision, microwave clocks involve transitions that depend on more fundamental constants and are therefore essential for constraining those dependencies.

Measurements made with the rubidium fountains (and other clocks at USNO) after the first 1.5 years of continuous operation resulted in the most precise test of the universality of the gravitational redshift and enabled constraints on the coupling of fundamental constants to gravity to be improved by on average a factor of four [7]. With 4.5 years of data, we expect to be able to improve upon these measurements. The highest precision comparison we can carry out is a comparison of rubidium and hydrogen, using our four fountains and many masers. Looking for an annual oscillation in 4.5 years of data enables most long-term maser drift to be taken into account without affecting the result. An important technique that was used in our original analysis was to determine baselines for EEP-violating signals among different pairs of masers, which enabled a bias for the EEP-violating signal to be assigned to each maser. This eliminated 'false positives' that otherwise would have appeared in inter-species comparisons. A new EEP analysis is underway and will hopefully be finalized in the near future.

The clocks in these measurements experience the same gravitational potential; using clocks at different locations on the Earth puts them at different solar potentials and enables a distinct test of EEP. While the size of the driving term – the variation in gravitational potential due to the Earth's rotation in the Sun's potential well – is less than that due to the Earth's elliptical orbit, the putative signal would arise as a daily oscillation. Experiments planned for space considered carrying out this type of test using common view satellite comparisons of clocks separated by intercontinental distances [8]. Terrestrial measurements using optical clocks connected via fiber networks are a promising avenue for this type of test. We have considered using clocks at USNO in Washington, DC and in Colorado Springs, CO, which are compared regularly via two-way satellite time and frequency transfer. These locations give a difference in gravitational potential of more than $10^{-13}$, which should lead to a limit on EEP violation for this type of test of about 3 orders of magnitude higher than the measurements with co-located clocks discussed above. The noise associated with satellite frequency transfer will likely limit the precision above this level, however.

## 5. Acknowledgements

We thank Paul Koppang and Jim Skinner for assistance with obtaining various types of clock data. Demetrios Matsakis has been involved with the study of EEP violation using clocks at different locations on the Earth.